\documentclass[onecolumn]{mn2e}  
\usepackage{mnras_cite}  
\usepackage[dvips]{graphicx}  

\usepackage{amsmath}
\newcommand{\nc}{\newcommand}   

\nc{\citealt}{\cite}
\nc{\de}{\delta}
\nc{\tISW}{\triangle_T^{ISW}}
\nc{\hn}{\hat{n}}
\nc{\bH}{\bar{H}}
\nc{\Ol}{\Om_{\Lambda}}

\nc{\ul}{\underline} \nc{\al}{\alpha} \nc{\g}{\gamma}
\nc{\Del}{\Delta} \nc{\e}{\textrm{e}} \nc{\eps}{\epsilon}
\nc{\lam}{\lambda} \nc{\Om}{\Omega} \nc{\Omm}{\Omega_m}
\nc{\Oml}{\Omega_\Lambda} \nc{\LCDM}{$\Lambda$CDM~} 
\nc{\ve}{\varepsilon} \nc{\mn}{{\mu\nu}} \nc{\vp}{\varphi}

\def\gsim{\; \raise0.3ex\hbox{$>$\kern-0.75em
\raise-1.1ex\hbox{$\sim$}}\; }

\nc{\Section}[2]{\section{#2}\label{#1}}   
\nc{\Bibitem}[1]{\bibitem{#1}}   
\nc{\Label}[1]{\label{#1}}   

\nc{\beq}[1]{\begin{equation}\label{#1}}     
\nc{\eeq}{\end{equation}}

\nc{\hq}{\hat{q}}
\nc{\hw}{\widehat{w}}

\def\ben{\begin{enumerate}}
\def\een{\end{enumerate}}
\def\bi{\begin{itemize}}
\def\ei{\end{itemize}}
\def\ee{\end{equation}}
\def\bea{\begin{eqnarray}}
\def\eea{\end{eqnarray}}

\def\mg{\big<}
\def\md{\big>}

\nc{\M}{\rm{M}}
\nc{\Msun}{\rm{M_{sun}/h}}
\nc{\Gpcc}{\rm{~ Gpc^3/h^3}}     
\nc{\Gpc}{\rm{~ Gpc/h}}     
\nc{\Mpc}{\rm{~ Mpc/h}}   
\nc{\vev}[1]{\langle #1 \rangle}   
   
\def\etal{{et al. }}   
   
\def\ltsima{$\; \buildrel < \over \sim \;$}   
\def\gtsima{$\; \buildrel > \over \sim \;$}   
\def\simlt{\lower.5ex\hbox{\ltsima}}   
\def\simgt{\lower.5ex\hbox{\gtsima}}   
\nc{\w}{$w(\theta)$\ }   
\nc{\ie}{i.e., }    
\nc{\eg}{e.g., }

\input epsf

\begin{document}   
   
\title[The onion universe]
{The onion universe: all sky lightcone simulations in spherical shells}

\author[Fosalba \etal]{Pablo Fosalba,
Enrique Gazta\~{n}aga,
Francisco J Castander \& Marc Manera \\ 
Institut de Ci\`encies de l'Espai, IEEC-CSIC, Campus UAB,
F. de Ci\`encies, Torre C5 par-2,  Barcelona 08193, Spain}

\twocolumn   
\maketitle 

\begin{abstract}

Galaxy surveys provide a large-scale view of the universe that typically
has a limited line-of-sight or redshift resolution.
The lack of radial accuracy in these surveys can be modelled by picturing the universe
as a set of concentric radial shells of finite width around the observer, i.e, an onion-like structure.  
We present a new N-body simulation with $2048^3$ particles developed at the 
Marenostrum supercomputer with the GADGET-2 code.
Using the lightcone output we build a set of angular maps
that mimic this onion-like representation of the universe.
The onion maps are a highly compressed version of the raw data
(\ie a factor $>1000$ smaller size for arcminute resolution maps) 
and they provide a new and powerful 
tool to exploit large scale structure observations.
We introduce two basic applications of these maps that are 
especially useful for constraning dark energy properties:
the baryon acoustic oscillations (BAO) in the galaxy power spectrum and 
all-sky maps of the weak lensing distortion.
In particular, from the matter density maps, we determine the smallest scale where linear
theory and the Gaussianity of the error analysis applies. 
Using the weak lensing maps, we measure
the convergence power spectra and compare it to halo fit predictions. We also 
discuss mass resolution effects and error determinations. 
As a further application, we compute the variance and higher-order moments of the maps. 
We show that sampling variance on
scales of few degrees is quite large, resulting in a significant 
(25\% at 10 arminute scales) bias in the variance.
We caution that current lensing surveys such as the COSMOS HST should
take into account this bias and extra sampling error in their clustering analyses and
inferred cosmological parameter constraints.
Finally, we test the importance of projection effects in the
weak lensing mass reconstruction. On the mean, the mass calibration
works well but it exhibits a large non-Gaussian scatter what could induce
a large bias in the recovered mass function.

\end{abstract}   
   

\section{Introduction}   

Upcoming astronomical surveys will gather many Terabytes of
unprecedented high quality data containing the relevant information to
answer key cosmological questions, ranging from the nature of the
initial conditions in the structure formation of the universe, how
galaxies and clusters form and evolve, or the properties of the so
called dark energy and theory of gravity on cosmological scales.

These datasets will pose a great challenge to the scientific community
to develop the appropriate data analysis tools to compress the
overwhelming raw data into a few numbers, eg a set of cosmological
parameters.  Simulating surveys, with their anticipated volume,
resolution and complexity, has become a standard tool to prepare the
scientific exploitation and to understand real astronomical
data. Thus, analyzing mock surveys suffers from similar limitations,
with the agravant that it requires a large number of simulations
to pin down statistical errors and explore cosmological parameter space.

Measuring redshifts for many galaxies is very costly (especially at
$z>1$), even for very large telescopes.  Thus, to explore the largest
scales with catalogs containing many millions of
galaxies typically requires a photometric approach to obtain galaxy
redshifts, as is the case for most of the upcoming or planned surveys
(such as DES, PAU, VHS, PanSTARRS, LSST, DUNE).  In these surveys
information in the radial direction is washed out on scales smaller
than the photometric error width. This limits the amount of
information that can be used for scientific analysis.

With this motivation, in this paper we develop a new approach to
building mock surveys, that we dub the ``onion universe'', which mimics the
tomographic structure of photometric surveys by decomposing the full
3D lightcone data structure into a set of concentric 2D all-sky maps
around the observer.  
We thus propose this method as a new and efficient tool to exploit 
upcoming large photometric surveys.
In particular, this new approach allows as to make the following main contributions
that we present in this paper: 

\begin{itemize}

\item {\it Data Compression:} 
for most of its cosmological applications, our approach provides an effectively
lossless method to compress simulated data by a factor $\sim 1000$ 
for arcminute resolution angular maps. 
This allows Terabyte-sized simulations containing tens of
billions of particles to be analyzed in a regular laptop.  

\item {\it BAOs in the angular power spectrum:}
using the set of dark-matter density shells resulting from our 
decomposition of the N-body lightcone data, we can straighforwardly   
study the BAOs in the angular clustering of dark-matter. 
As an application of this, we assess the limit of applicability of linear theory
(and Gaussianity of the error) as a function of redshift from an angular power spectrum analysis.

\item {\it All-sky weak lensing maps:}
our method 
provides a straightforward way to simulate all-sky maps of tracers of the large-scale structure 
in the light-cone from weighted combinations of 2D density maps.
The main application of the ``onion universe'' method presented in this paper 
is the development, for the first time, 
of an adequate all-sky simulated weak lensing map with fine angular resolution. 
Our map is validated through a comparison with theoretical predictions for 
the power spectrum over 3 decades in angular scales. 
We use this mock map to investigate effects of sampling bias in current weak lensing surveys
from higher-order moments of the convergence field, and discuss the potential and limitations
of using weak lensing as a cluster mass calibrator. 

\end{itemize}

Weak gravitational lensing by the large-scale structure of the
universe probes density fluctuations in a wide dynamical range, from
linear to highly non-linear scales. Current lensing observations (eg
\pcite{Bacon01,RRG02,JJBD06,COSMOS}) can only sample the smaller
scales (ie smaller than a degree) which are dominated by non-linear
fluctuations. Since there is no accurate analytic description of the
dark matter density field in the non-linear regime it is thus
necessary to resort to numerical simulations to make accurate
predictions of the weak lensing distortions.

Simulations of weak gravitational lensing are based on implementations
of the ray tracing technique on N-body simulations (eg see
\pcite{WCO98,JSW00,VW03}, and references therein).  In this approach
light rays are propagated from the observer to the source by computing
the distortion and magnification effects from multiple (from tens to a
hundred) of equally spaced projected-mass lens planes.  This approach
has proven to be successful in measuring the lensing power spectrum
which was found to be in agreement with the Born and Limber
approximations (eg \pcite{JSW00,VW03}). In addition, the lensing
higher order moments induced by density fluctuations in the non-linear
regime can also be estimated (eg \pcite{JSW00,WH00,VW03,WV04}) and
they turn out to be consistent with perturbation theory results on the
largest scales and analytic fits on intermediate scales 
(eg \pcite{GB98,W01}).  In particular, measurements of the variance
and skewness of the lensing maps can be used to constrain the
amplitude of matter fluctuations, $\sigma_8$, and matter density,
$\Omega_m$ \cite{BWM97,JS97}.

However, covering a wide enough dynamical range (from Mpc to Gpc
scales) is prohibitive with current implementations of ray tracing
techniques and therefore simulations so far have focused on small
patches of the sky (i.e few square degrees), comparable to the areas
covered by current surveys (such as GEMS, COSMOS HST, CFHTLS), well in
the non-linear regime, where most of the lensing signal is expected to
come from.  Moreover, as we will show below, statistical measurements
in small volume simulated surveys may be significantly affected by
sampling bias \cite{HG98} and cosmic variance errors are largely
enhanced by non-Gaussianity \cite{Scocci99,Semboloni07}.  For a review on the
weak lensing formalism, simulations and observations see \scite{BS01},
\scite{Refregier03} and references therein.

Previous work on simulating observational data in terms of lightcone surveys
has concentrated on galaxy mocks (see e.g, \pcite{Blaizot05,KW07,FR07} and references therein), 
where finite simulation volume, redshift discreteness, and cosmic
variance were carefully addressed.

This paper is organized as follows: Section 2 presents the
simulations, the onion maps and its power spectrum. In Section 3, we
introduce the new method to build the weak lensing maps from the onion
shells and present several tests to validate the method. We also 
discuss different error estimates based on the convergence maps
and a comparison with the halo-fit prediction. Section 4 is devoted to
the study of non-Gaussianity, mass reconstruction and mass function.
Finally, in Section 5 we summarize our main results and conclusions.

\section{Onion maps}   
\label{sec:N-body}

\begin{table}{}
\caption{MICE N-body simulations used in this paper.}
\begin{tabular}{|c|c|c|}
\hline
$L_{box}$ & $N_{par}$ & particle mass \\
$\Mpc$ & number & $10^{10} M_{sun}/h$  \\
\hline
$3072$ & $512^3$ &  $1510$\\
$3072$ & $1024^3$ &  $190$\\
$3072$ & $2048^3$ &  $24$\\
\hline
\label{table1}
\end{tabular}
\end{table}

We have developed a set of large N-body simulations with Gadget-2 \cite{Springel05}
on the Marenostrum supercomputer at BSC \footnote{Barcelona Supercomputing Center, www.bsc.es}.
We shall name them MICE (MareNostrum - Instituto de Ciencias del Espacio) 
simulations hereafter
\footnote{Projected matter density and weak lensing maps 
from the MICE simulations are publicly available at http://www.ice.cat/mice}.  
In this paper we focus on a simulation
run with $2048^3$ dark-matter particles in a box-size of $L_{box}= 3072 \Mpc$,
and assume a flat concordance LCDM model with $\Omega_m=0.25$,
$\Omega_\Lambda =0.75$, $\Omega_b=0.044$, $n_s=0.95$, $\sigma_8=0.8$
and $h=0.7$. The resulting particle mass is $\M = 2.34 \times 10^{11} \Msun$ and  
the softening length used is 50 \rm{Kpc/h}. Thus our
simulation has a dynamic range close to five orders of magnitude.  
We start our run at $z_i = 50$ displacing particles using the Zeldovich dynamics.
The MICE simulation has a similar number of particles to the Millenium
simulation \cite{millenium} but $6^3=216$ times more volume (and
corresponding larger particle mass). This makes the MICE simulation
more adequate than the Millenium for very large scale statistical
analyses, such as the search of the baryon acoustic scale (see below)
and the study of very long distance effects (such as gravitational
lensing). The drawback is the limited resolution that is required
to study smaller scales and substructure in galaxy size halos. 
In terms of volume, our simulation is similar to the Hubble Volume 
\cite{Evrard02}, but it has 8 times better mass resolution.
Given that we have a relatively large particle mass, we have also used some MICE
runs with different number of particles, given in Table \ref{table1},
to test resolution effects.  
We have also varied the box size,
ranging from $L_{box}= 768 \Mpc$ to $L_{box}= 7680 \Mpc$, to explore
volume effects. Results from this analysis will be discussed in detail elsewhere.

We have built a lightcone output of this simulation from
$\sim 200$ comoving outputs which are separated by constant spacing
in cosmic time ( $\approx 70$ Myr ). The ligtcone has been constructed
in spherical concentric shells, each one getting its particles
from one of the comoving outputs. The distance from the center of
the set of spherical shells, 
where the observer is, to the mean radial
distance of each shell is given by the corresponding redshift of each
comoving output. In every shell the dark matter particles are moved
using their peculiar velocities in order to extrapolate them into
their lightcone positions. We have allowed a spherical shell buffer
to take into acount the particles that cross shell boundaries when
moved.

The lightcone extends out to $z_{LC} \simeq 6$ 
(\ie a comoving distance $r_{LC} \simeq 6 \Gpc$ from the observer) 
by replicating twice the parent simulation
with $L_{box}$ size along each cartesian direction and applying periodic
boundary conditions. Note that a radial comoving distance of $3 \Gpc$ corresponds to 
$z \simeq 1.4$, where cosmic evolution plays a significant role for an observer at 
$z = 0$, so we do not expect this periodic repetition to have much impact on our
analysis. We plan to explore this issue further in future work by comparing
results of different $L_{box}$. Preliminary results indicate that
$L_{box}= 3072 \Mpc$ is large enough for most applications we have
explored.

We have done several tests with the simulation output to make sure
that basic clustering statistics, such as the power spectrum, the halo mass function and the higher
order correlations, are consistent with previous results. 
As shown below, we have measured the baryon acoustic oscillations (BAO) imprinted 
in the matter distribution at different redshifts or
onion shells and found good agreement with other analyses.  More
details will be presented elsewhere.

\subsection{Redshift shells and BAO}

\begin{figure}
\includegraphics[width = 3.3in]{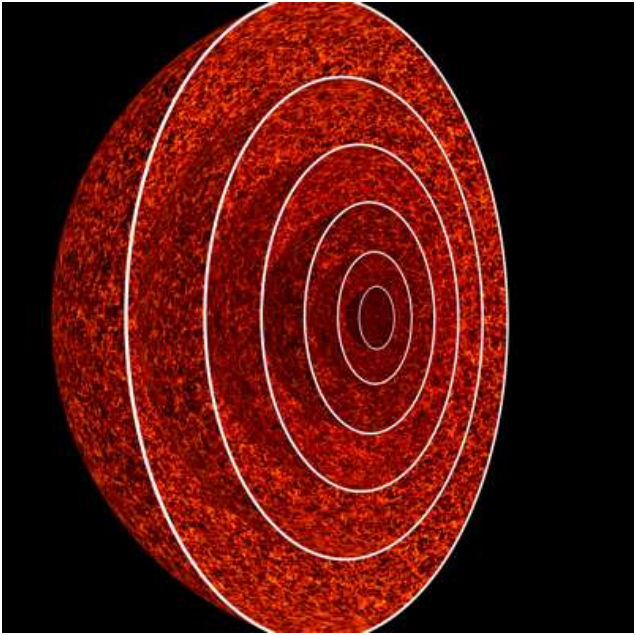}
\caption{The onion universe: a decomposition  of the lightcone that mimics the data structure in photometric galaxy surveys. The simulated universe is rendered as a discrete set of projected matter density shperical shells in the lightcone around the observer, i.e, at the center of the concentric spheres. 2D spherical shells are equally spaced in comoving time and pixelized using the Healpix tesselation of chosen angular resolution. For clarity, in this figure we only show one of the hemispheres (i.e half the onion universe) for several of the lowest redshift shells.}
\label{fig:onion}
\end{figure}

We discretize the lightcone output of the simulation in concentric
spherical shells of width $dz$ given by a constant spacing in cosmic
time.  We use a spacing of $\approx 70$ Myr to match the width used
for the shells in the lightcone construction described above. 
This corresponds to a $dz$ that varies from
$dz \simeq 0.005$ at low $z$ to $dz \simeq 0.025$ at $z=1.4$ (which
corresponds to a width of $\simeq 16 \Mpc$ to $35 \Mpc$).  This is
probably enough for most large scale applications. Each onion shell is
stored as a number density pixel map of given resolution in the
Healpix format \cite{GHW99}. Here we use maps with $N_{side}=2048$, which pixelices the
sky with $12 N_{side}^2 \approx 50$ million cells of size $\theta_{pix} \simeq 1.7$ arcmin size.
The resulting onion universe decomposition of the parent simulation is 
shown in Fig.\ref{fig:onion}.

\begin{figure}
	{\centering{\epsfysize=9.cm \epsfbox{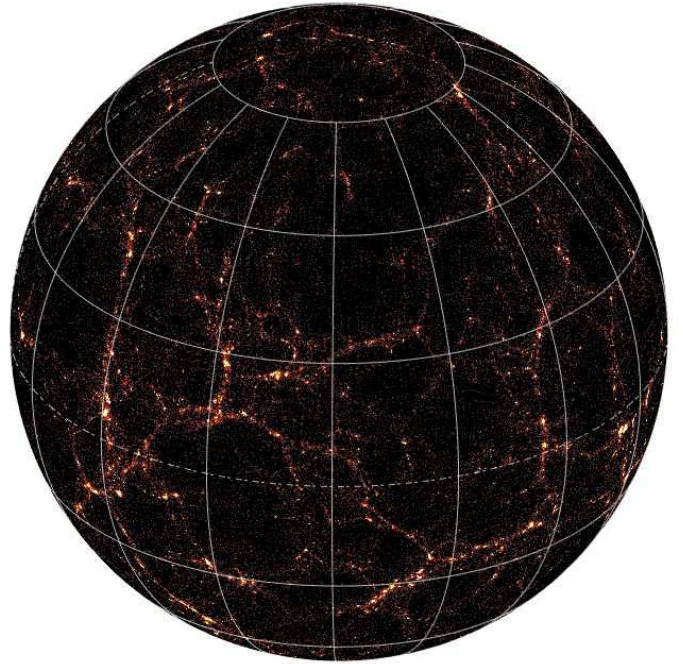}}
	}
\caption{Onion shell density map 
at $z\simeq 0.036$ (this corresponds to a comoving distance of $r=108 \pm 8 \Mpc$)}
\label{fig:maps1}
\end{figure}

\begin{figure}
	{\centering{\epsfysize=9.cm \epsfbox{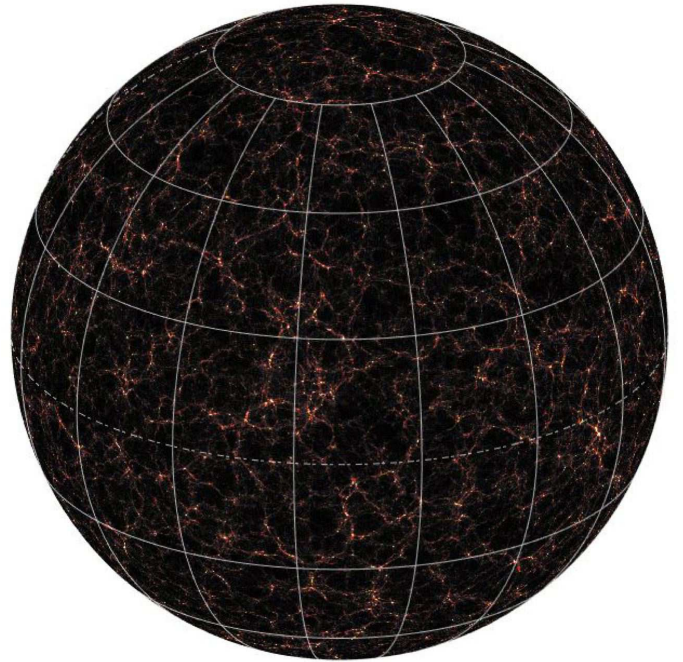}}
	}
\caption{Onion shell density map 
at $z\simeq 0.15$ (comoving distance $r=439 \pm 9 \Mpc$)}
\label{fig:maps3}
\end{figure}

\begin{figure}
	{\centering{\epsfysize=9.cm \epsfbox{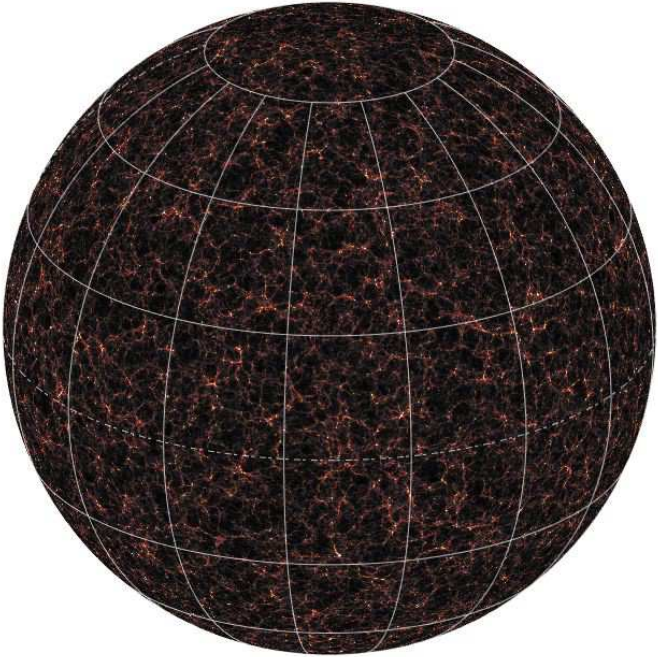}}
	}
\caption{Onion shell density map 
at $z\simeq 0.30$ (comoving distance $r=866 \pm 10 \Mpc$)}
\label{fig:maps4}
\end{figure}

\begin{figure}
	{\centering{\epsfysize=8.5cm \epsfbox{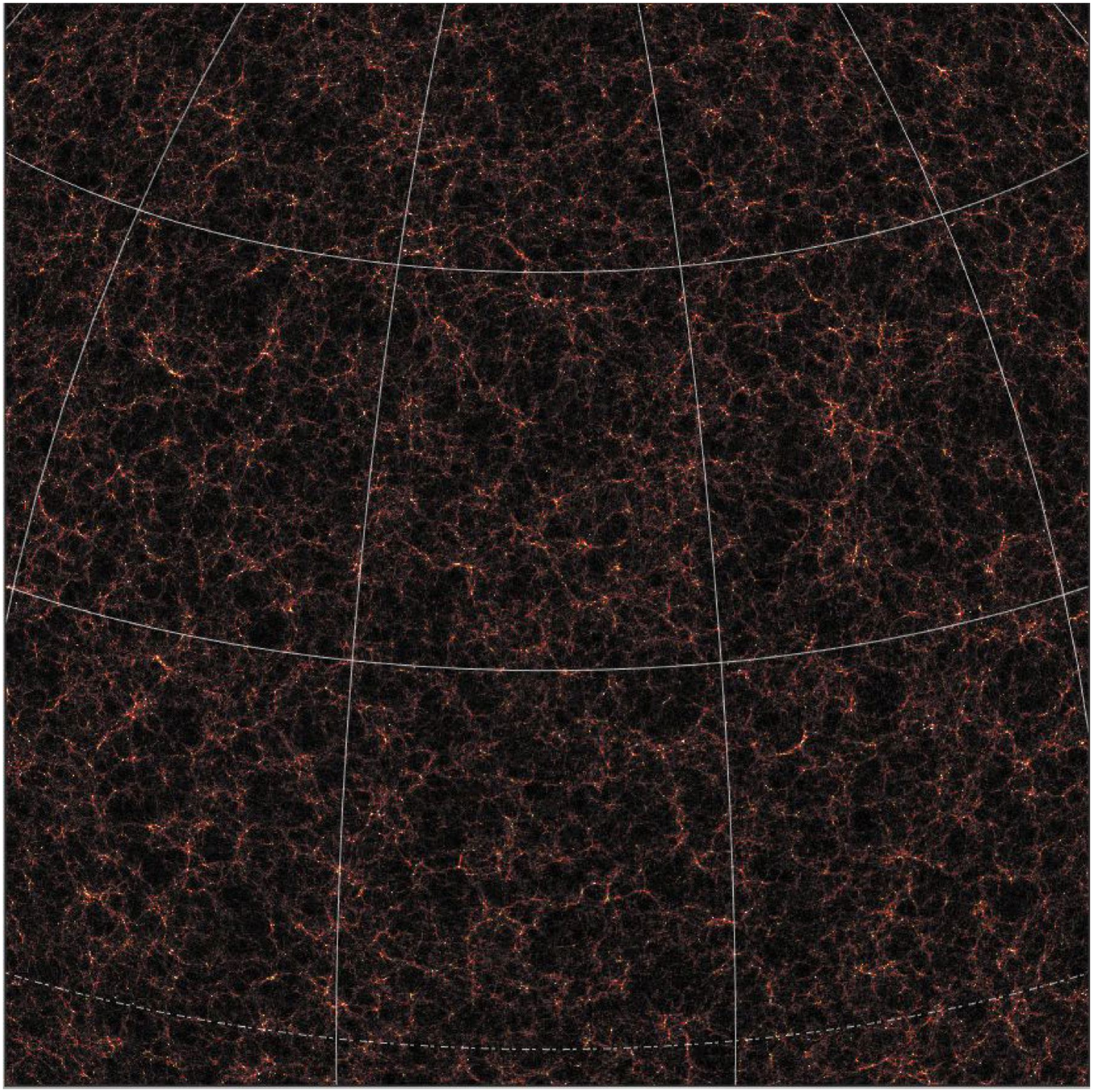}}
	}
\caption{Zoom over a shell at $z\simeq 0.60$ (comoving distance $r=1589 \pm 12 \Mpc$).}
\label{fig:maps2}
\end{figure}

Fig.\ref{fig:maps1}-\ref{fig:maps2} show four spherical shells of the onion universe
corresponding to the projected matter density distribution 
for four different redshifts (with brighter colours representing higher number
of particles per pixel in a log scale). It is clearly visible a 
characteristic $\sim 100 \Mpc$ cell of filamentary structure 
(the so-call Cosmic Web, ie \pcite{BKP96}). The cell-size 
shrinks to smaller angular scales and smaller
particle number density contrast as we move to larger redshifts.  
By $z=0.6$, there are already close to a
thousand of these $100 \Mpc$ cells in this single onion shell (\ie Fig.\ref{fig:maps2}).  
It is precisely around this $\simeq 100 \Mpc$ scale that future
surveys will aim at measuring the  
baryon acoustic oscillations scale, $r_{BAO}$.  The relative error involved in
measuring $r_{BAO}$ is roughly proportional to the inverse of the
square root of the number of independent $r_{BAO}$ cells: \beq{eq:bao}
\Delta_{BAO} \equiv {\Delta r_{BAO}\over{r_{BAO}}} \simeq
\left({r_{BAO}^3\over{V}} \right)^{1/2} \eeq where $V$ is the sampled
volume, and we have assumed Gaussian errors (with negligible
shot-noise) over the first two BAO wiggles (see also \pcite{ABFL08}). 
Thus, for the onion shell at $z=0.6$ we estimate $\Delta_{BAO} \simeq
1/\sqrt{1000} \simeq 3\%$.  According to this rule of thumb, we can
get to $0.6\%$ relative error in measuring $r_{BAO}$ using the whole
MICE simulation volume, as compared to $9\%$ with the Millenium
simulation.

\subsection{Compression factor}

To build the light-cone with sufficient accuracy, we have used 200
comoving simulation outputs. Each output takes 250 Gbytes, so the
total storage required is about 49 Terabytes.  If we match the spatial
width of the onion shells (as we have done) to the time lag between
the outputs that are used to build the light-cone we will have
equivalent information for applications that do not require angular or
redshift resolution better than that projected onto the pixel maps. We have
produced 200 such Healpix maps, each occupies 201 Megabyte, which
represents a total of 39 Gigabytes.  Thus, there is total compression
factor of about 1300 when using the pixel maps instead of the full
comoving outputs.  Nowadays, it is practically impossible to manage
and share 49 Terabytes of data in a public network: it will take
more than a year to transfer these data over a 10 Mbit/sec
connection. On the other hand, 39 Gbytes fits into a laptop and can be
shared in a matter of hours. There are applications for which we might
need the full MICE output information, but in many cases the pixel
maps will be very useful, specially when we compare to observations in
photometric surveys, as we show next.

\subsection{Angular Power spectrum}

\begin{figure}
	{\centering{\epsfysize=8.5cm \epsfbox{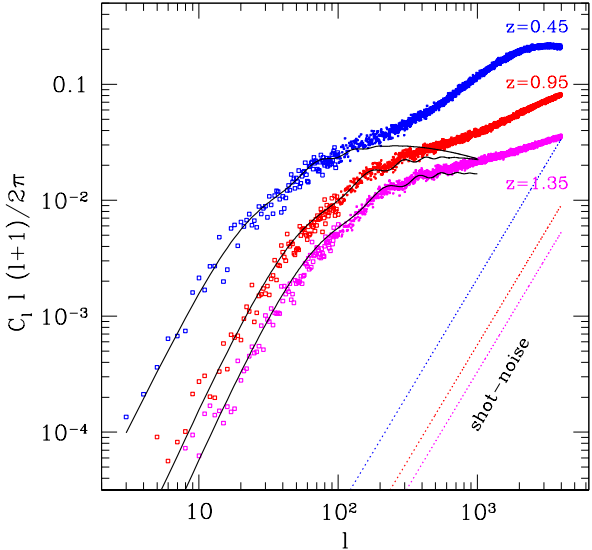}}
	}
\caption{Angular power spectrum estimated from combinations of onion
maps (symbols) of width $dz=0.10$ and mean redshifts of $z=0.45$
(top), $z=0.95$ (middle) and $z=1.35$ (bottom). The continuous line shows
the linear theory prediction for each redshift. The dotted line shows the
shot-noise contribution in each case.}
\label{clgg}
\end{figure}

Fig.\ref{clgg} shows the total power per log multipole interval $P_l
\equiv l(l+1)~C_l/2\pi $ measured from the all-sky onion maps. We have
combined thin onion maps into $dz=0.1$ slices to match the
photo-z error expected for a survey such as the photometric
SDSS \cite{sdssDR4} or DES \cite{DESWP}\footnote{We take the mean
depth of the survey to be $z \simeq 0.7$, but this has a very small
effect over the maps because the selection function is normalized to
be unity in each slice}. Symbols in the figure correspond to the $P_l$
estimation in maps with redshift ranges (from top to bottom):
$z=0.4-0.5$, $0.9-1.0$ and $1.4-1.5$.

As expected, the amplitude of the fluctuations decreases with the
depth of the slice. The BAO wiggles (around $l \simeq 80-300$), which
are clearly visible, also move to smaller angular scales (higher
multipoles).  The scatter within multipole bins (\ie the ``intra-bin'' variance) 
estimates the sampling variance for band limited measurements \cite{Fosalba04}, although this 
method requires appropriate calibration with simulations. 
In Section 3 we compare a simple implementation of the intra-bin variance against other 
more standard error estimators. 
Shot-noise (shown as a dotted line) does
not affect the $C_l$ estimation within the range of scales shown.

Linear theory predictions are shown by the continuous lines.
Simulations match well the linear prediction up to scales of around
the first BAO wiggle. On smaller scales (larger multipoles) non-linear
effects become important. This effect is larger at low
redshift, as expected. For the smallest redshift bin shown
($z=0.4-0.5$), there is a flattening of the power on scales $l>2000$
indicating the virialization of structures.

\begin{figure}
	{\centering{\epsfysize=8.5cm \epsfbox{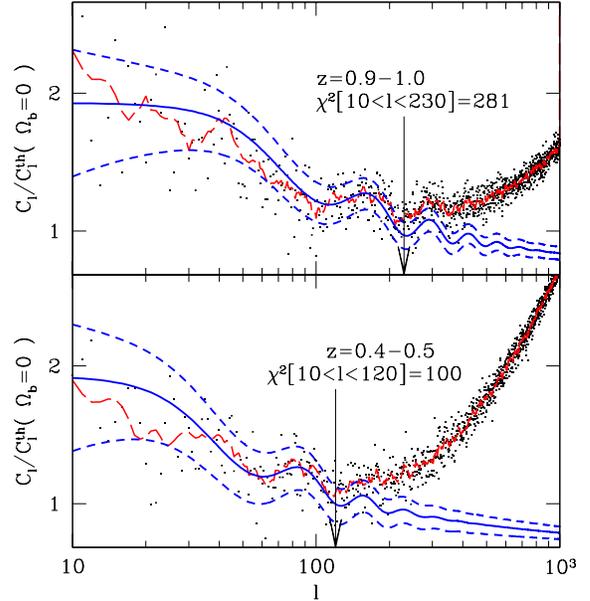}}
	}
\caption{Angular power spectrum normalized to a linear model without
baryons. The continuous line corresponds to the linear theory prediction
(with baryons) while the dashed lines marks the 1-sigma Gaussian errorbars.
The dots correspond to measurements in the simulated maps.
The long-dashed (red) line just shows a smooth version of the dots.}
\label{nclggz}
\end{figure}

Fig.\ref{nclggz} highlights the BAOs in the $C_l$'s 
by normalizing to the linear theory prediction without baryons (\ie same
cosmological parameters but $\Omega_b=0$). The first prominent step at
$l_h=10$ corresponds to the horizon scale at the matter-radiation
transition. The first BAO is found at around $l_1=80$ for
$z=0.4-0.5$ and $l_1=160$ for $z=0.9-1.0$.
These multipoles correspond to the projected scale for the first peak 
in the 3D power spectrum, $k_1 \simeq l_1/r(z) \approx 0.07$, where r(z) is the comoving 
distance to the onion slice and the approximation is valid 
for small angular scales (\ie we employ the Limber approximation).

On scales where linear theory is valid and for $l>10$ it is usually
assumed that the $C_l$ amplitudes follow a Gaussian distribution with
sampling errors given by the Gaussian prediction $\sigma^2_G = 2
C_l^2/(l+1)$. We can test this hypothesis with our maps by calculating
the Gaussian $\chi^2$: 
\beq{chi2} \chi^2 = \sum_{l=10}^{l=lmax}
{(C_l^m-C_l^t)^2\over{\sigma^2_G}} \eeq 
where $C_l^m$ are the measured values and $C_l^t$ are the linear
theory predictions. 
We find $\chi^2 = 281$ and $\chi^2=100$ for
$z=0.95$ and $z=0.45$ respectively. 
We use $lmax=230$ and $lmax=120$ to include the whole first BAO
wiggle. These values are larger than expected for a sum of Gaussian
variables for the $z=0.95$ case, and are fine for $z=0.45$.  The
probabilities of this to happen are $P[\chi^2]=70\%$ for $z=0.45$ and
$P[\chi^2]=0.26\%$ for $z=0.95$. The later probability increases to
$P[\chi^2]=12\%$ for $z=0.95$ if we use $lmax=177$ (half the way
through the first BAO wiggle).  This indicates that even on these very
large scales and early times there is correlation between different
modes and one should be careful when doing precision forecasts using
Gaussian errors. The correlation between bins is weaker if we bin the
data in multipoles (see below). However, the problem seems more critical at high
redshift where the BAO wiggle is better sampled and there are
larger projection effects. A more detailed analysis of this effect
requires more realizations and will be presented elsewhere.

\section{Convergence maps}

\begin{figure*}
	{\centering{\epsfysize=9cm \epsfbox{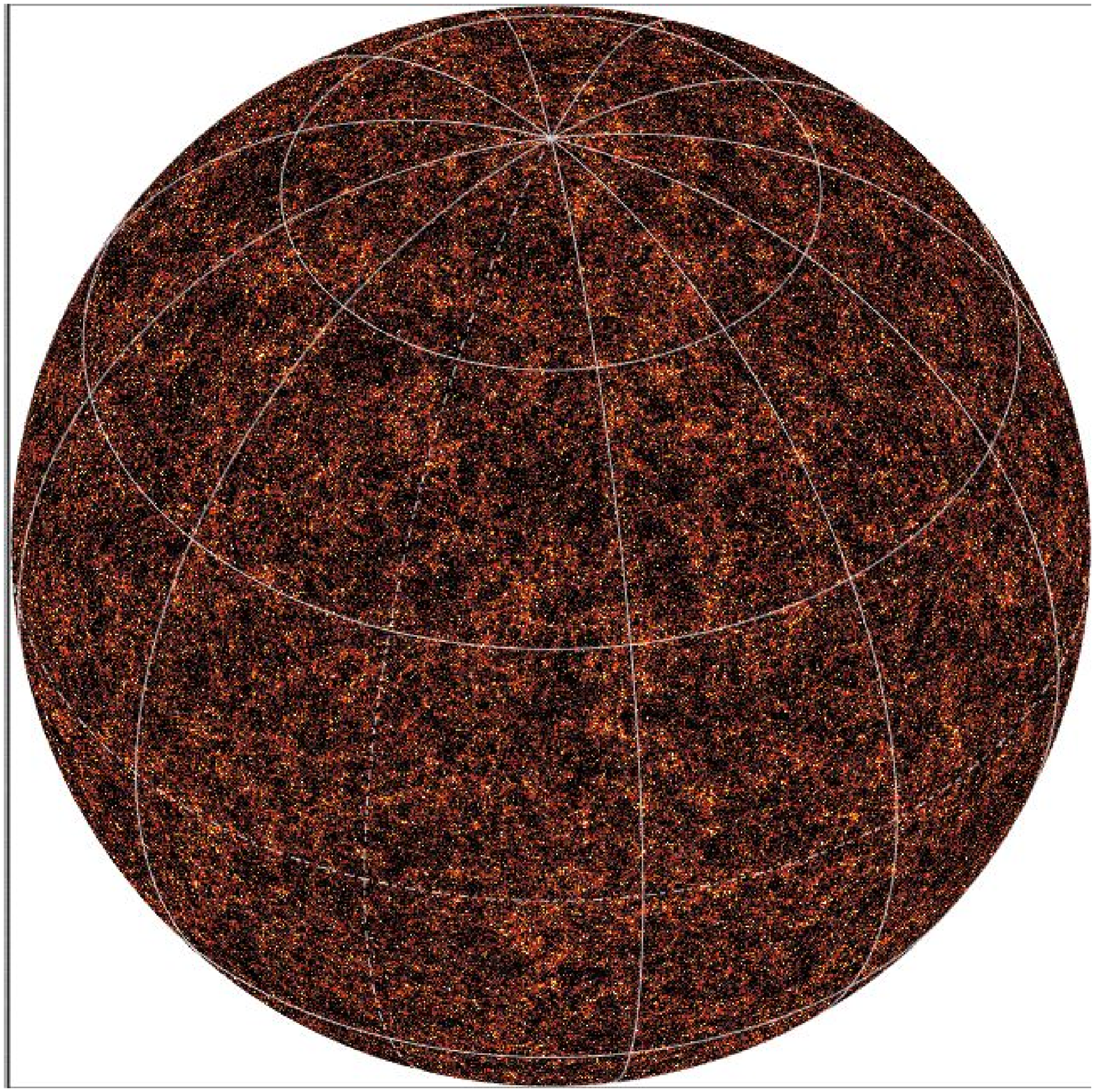}}
	\centering{\epsfysize=9cm \epsfbox{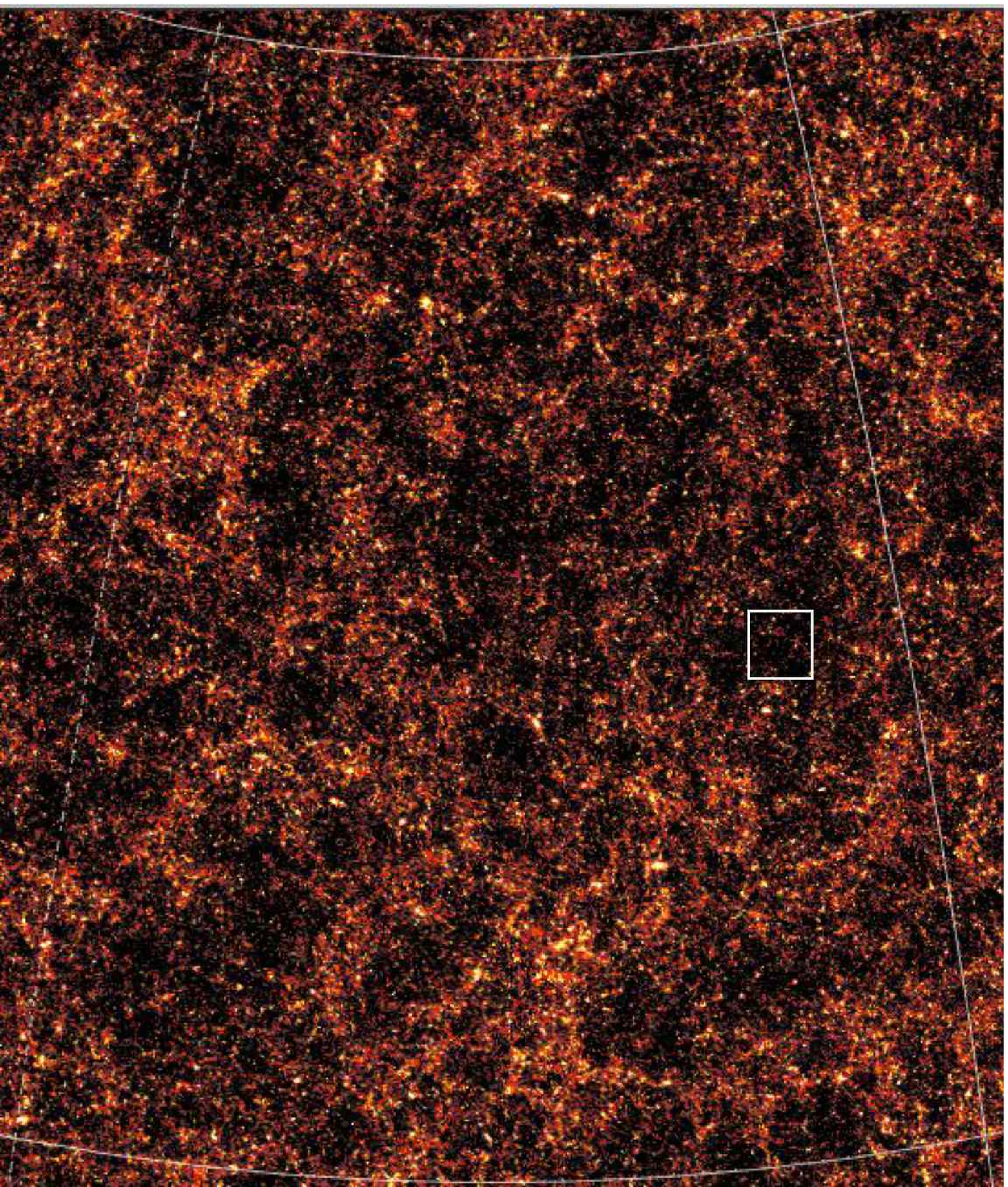}}
	}
\caption{Left: All-sky convergence map with sources at $z_s=1$. 
Right: A $15 \times 15$ degrees zoom in the central region. The small white square
shows an area of 1.6 $\rm{deg^2}$, comparable to the COSMOS HST survey area.}
\label{kmaps}
\end{figure*}

A basic quantity in weak lensing is the convergence field. In
the so-called Born approximation, one integrates the lensing
distortion over the unperturbed photon paths which is a good
approximation for most applications (eg see
\pcite{CH02,BS01,VW03,Refregier03}).  In this approximation, the
convergence is just a weighted projected surface density: \beq{kappa}
\kappa(\theta) = {3H_0^2\Omega_m\over{2c^2}}~\int ~dr~\delta(r,\theta)
{(r_s-r)r\over{r_s~a}} \eeq where $\delta$ is the 3D matter density at
radial distance $r$ and angular position $\theta$ (which is here a 2D
vector) and $r_s$ is the radial position of the lensing
sources.\footnote{Without loss of generality we will assume on
writing equations that we live in a flat universe and that all
lensing sources are at a fixed redshift $z_s$, or radial distance
$r_s$. It is straightforward to generalize this to an arbitrary
distribution of sources, eg \cite{BS01,Refregier03}.}  Out of $\kappa$
we can get all other quantities of interest such as magnification,
demagnification, shear, projected potential or deflection angle (see
\pcite{BS01,Refregier03} for a review).  Some of these relations are
not local (they involve derivatives or integrals) but for an all-sky
map, all these quantities are trivially related. In harmonic (or 2D Fourier)
space, we can transform a convergence map into a gravitational potential or 
shear map by just multiplying the harmonic (or Fourier) $\kappa$ amplitudes by the
appropriate combination of multipole or wave numbers. However, in the
observable universe the boundaries of a real survey complicate these
transformations.

We will build our convergence map by just adding the onion slices from the
simulation with the appropriate lensing weight. This can be done as
follows (see also \pcite{GB98}):

\beq{kappasum}
\kappa(i) = {3H_0^2\Omega_m\over{2c^2}}~\sum_j ~\delta(i,j)~{(r_s-r_j)r_j\over{r_s a_j}}~dr_j 
\eeq
where $i$ indicates a pixel position in the sky and $j$ a radial bin
(at distance $r_j$ of width $dr_j$) into which we have sliced 
the simulation as described in the previous section. 
If we indicate by $N_{ij}$ the number of particles in pixel $i$ 
from onion slice $j$, we have:

\beq{deltaij}
\delta(i,j)={\rho(i,j)\over{\bar{\rho}}}-1
\eeq

where $\bar{\rho}= <\rho(i,j)>$ and

\beq{rhoij}
\rho(i,j) = {N_{ij}\over{dV_j}} = {N_{ij}\over{\Delta\Omega ~r_j^2~ dr_j}}
\eeq
where $\Delta\Omega$ is the area of each pixel. 

Fig.\ref{kmaps} shows images of the resulting maps for lensing sources
at $z_s=1$. Note how despite the large volume projected there is still
considerable structure in these maps. In particular, on scales
of a few square degrees there is a large variation from place to place
in the maps. This indicates that sampling variance is important.
Current weak lensing surveys, such as COSMOS \cite{COSMOS}, or lensing
simulations, only expand scales of the order of a few square degrees.  Our
convergence maps show that sampling variance is quite large on such
small scales and it is unlikely that current data could represent a
fair sample of the universe.  In section 4 and in the Conclusions we
will show some more quantitative consequences of this note.

\subsection{Validation of the map}

We can validate the convergence map by comparing its power spectrum and
higher orders to theoretical expectations.  First we will focus on a
comparison with linear theory and a consistency test for the angular
power spectrum.

Based on its definition in Eq.\ref{kappa} we would expect the angular
power spectrum of the convergence to yield:

\beq{clkappa}
C_l(\kappa) =
{9H_0^4\Omega_m^2\over{4c^4}}~\int ~dr~ P(k,z) 
{(r_s-r)^2\over{r_s^2~a^2}}
\eeq
where $P(k,z)$ is the 3D density power spectrum
in the simulation at redshift $z$ (corresponding to the radial
coordinate $r=r(z)$ in the integral) evaluated at $k=l/r$ 
in the small angle (Limber) approximation, valid for $l>10$ within 
a few percent accuracy (see e.g, \pcite{VW03}). In terms of discrete onion 
shells, this  translates into:
\beq{clkappasum}
C_l(\kappa) =
{9H_0^4\Omega_m^2\over{4c^4}}~\sum_j
 ~dr_j~ P(l/r_j,z_j) {(r_s-r_j)^2\over{r_s^2~a_j^2}}
\eeq

\begin{figure}
	{\centering{\epsfysize=8.cm \epsfbox{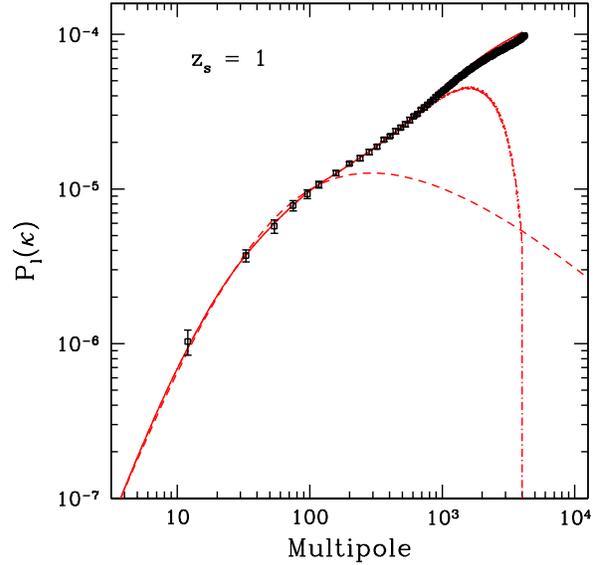}}
	}
\caption{Angular power spectrum in the convergence maps (symbols with errorbars)
as compared to linear theory (dashed line) and predictions from 
the full measured 3D power spectrum in Eq.\ref{clkappa} (line that goes
through the symbols). The dot-dashed line uses the same prediction with
 the 3D power spectrum measured
only using a $2048^3$ particle-mesh (PM) grid.}
\label{fig:clcheck}
\end{figure}

Fig.\ref{fig:clcheck} shows a comparison of the angular power spectrum
in the above prediction (continuous line through the symbols) to the
power spectrum directly measured from the convergence maps (symbols with
errorbars). The power spectrum has been binned in adjacent multipoles
with bin-width $\Delta l=20$.  The errorbars indicate the scatter of power
within a bin.  For the prediction in Eq.\ref{clkappasum} we have used
the actual 3D power spectrum measured from all particles in the
comoving outputs of the MICE simulation at the corresponding
redshifts. The dot-dashed line uses the density in a PM grid of
$2048^3$ to estimate the same $P(k,z)$. In both cases, this is just an
approximation because $P(k,z)$ should be the power spectrum in the
lightcone. But the difference is small because the redshift shell is
quite narrow. It is also an approximation in the sense that $P(k,z)$
in the whole box could be different to the power in a particular
onion shell (redshift bin), due to sampling variance.

The measured spectrum agrees with the linear prediction on the largest
scales ($l<100$), as expected. The agreement in shape and amplitude
with the prediction validates the convergence maps on the largest scales.
On all scales the agreement with Eq.\ref{clkappasum} is excellent,
indicating that the way we have built the convergence maps is a good
approximation to the true map. It also indicates that statistically,
inhomogeneities in the radial direction do not affect the projection,
which was assumed to get to Eq.\ref{clkappasum}.

The deviations at $l>1000$ when using the PM grid $P(k,z)$ (dot-dashed
line) show that the power spectrum on those scales come from structure
on scales smaller than the cell size used for the PM grid  $\simeq 1.5 \Mpc$. Thus, one
needs higher resolution PM simulations (or treePM algorithms for a given PM grid size) to
model the larger multipoles.

\begin{figure}
	{\centering{\epsfysize=8.cm \epsfbox{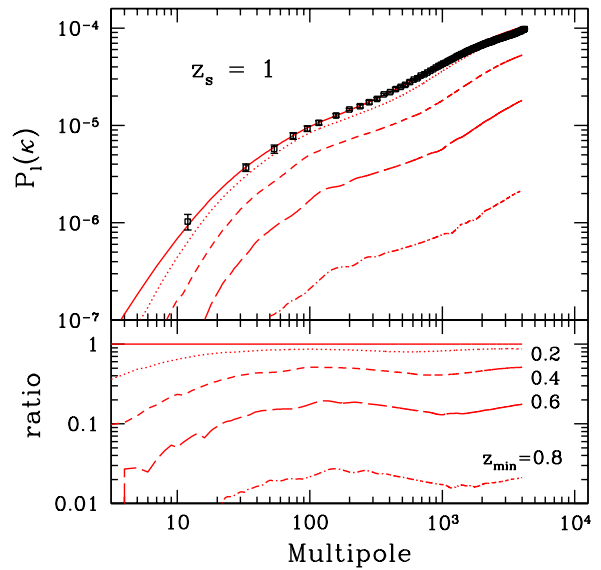}}
	}
\caption{The top panel shows as dot-dashed, long-dashed, short-dashed and dotted
lines the cumulative contributions to the sum in Eq.\ref{clkappasum}
starting from $z=z_{min}=0.8,0.6,0.5$ and $0.2$ 
respectively (as labeled in the figure) and integrated to $z=z_s=1$. 
The bottom panel shows the ratio of these quantities to the total contribution.}
\label{fig:clcheck2}
\end{figure}

In Fig.\ref{fig:clcheck2} we show the cumulative contribution in
Eq.\ref{clkappasum} from $z=z_{min}$ to $z=z_s$. For multipoles
$l>100$ the relative contributions are quite flat with $2\%$, $20\%$,
$50\%$ and $85\%$ contribution coming for sources at $z_{min}>0.8$, $z_{min}>0.6$,
$z_{min}>0.4$ and $z_{min}>0.2$, respectively.  All redshifts seem to
add power in the non-linear multipoles $l>1000$.  Note how the lowest
multipoles have a contribution $>50\%$ from $z<0.2$, due to local
structures.

\subsection{Error comparison}

The convergence map with sources at $z_s = 1$ shown in Fig.\ref{kmaps}, 
nominally covers a volume, $V_n\simeq 4/3\pi r(z_s=1)^3 \simeq 58 \Gpcc$ with 
$r(z_s=1) \simeq 2400 \Mpc$, which is twice the volume of the parent MICE simulation,
$V_p=L_{box}^3= (3072 \Mpc)^3 \simeq 29 \Gpcc$.
This is possible because we have replicated twice the box in each cartesian axis to
get to $z\simeq 6$ (see also Section 2). However, note that to reach to $z_s=1$ each sky octant
(1/8 of the full sky) is truly independent in the way we have built
the lightcone. Firstly because each sky octant uses the same replica but with the
observer placed in a different box position, so the same structures
are seen differently (\ie they are seen from a different angle and distance).  
Secondly, because of the shape of the lensing efficiency, the
shells closer to $z_s=1$ (where the shell-volume is larger) give a negligible
weak lensing signal. We can define an effective volume $V_e$ sampled
by the convergence map as the volume weighted by the lensing efficiency
function (renormalized to be one at the peak efficiency). We find that
in fact $V_e \simeq V_n/2 \simeq 29 \Gpcc$. We thus conclude that the convergence map to
$z_s=1$ samples well the full parent volume without significant repetition.

We can therefore split the sky in 10 equal area disjoint regions (a
partition each $10\%$ of the sky) and measure $C_l$ in each of the 10
regions using a fast 2-point estimator, SpICE \cite{spice,spice2}.  
We use the variance in the
$C_l$ from each region to have a direct estimate of the errors for a
survey that covers a fraction $f_{sky}=0.1$ of the sky.  We call this
the ``sub-sample'' (SUB) error. To avoid the covariance between bins, we
follow \cite{Cabre07} and bin adjacent $C_l$ estimations in bins of with $\Delta l=20$ for $l<100$ and 
$\Delta l=40$ for $l>100$. These binwidhts render the covaraince matrix 
effectively diagonal for the binned $C_l$'s  (see \pcite{Cabre07} for details).
We can also use a variance estimator based on 
the rms dispersion of individual $C_l$ amplitudes within in a given bin to estimate the
error in that bin. We shall call this ``intra-bin'' (IB) error estimation \cite{Fosalba04}. 
For Gaussian distributed amplitudes this is a good
estimate, because there is no correlation between adjacent multipoles.

We will compare the above errors with the traditional Gaussian estimate
of the sampling variance (SV):
\beq{sigmaG}
\sigma_{G} = C_l~\sqrt{2\over{f_{sky}~\Delta l~(2l+1)}}.
\eeq
where $\Delta l=N_{bin}$ is the number of multipoles in each bin.

\begin{figure}
	{\centering{\epsfysize=7.5cm \epsfbox{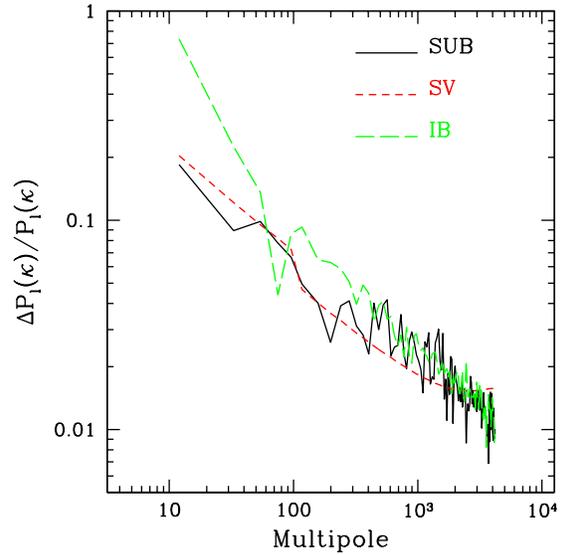}}
	}
\caption{Comparison of relative errors in 
$C_l(\kappa)$ for $10\%$ of the sky:
a) variance from 10 subsamples in all-sky convergence map (continuous line, SUB) b)
sampling variance from Gaussian statistics (dashed line, SV) and c) intra-bin
variance (long-dashed, IB).}
\label{fig:cle}
\end{figure}

Fig.\ref{fig:cle} compares the different estimates for the relative
errors in $C_l(\kappa)$. On scales $l>1000$ there is a good general
agreement. SV errors seem to
underestimate SUB errors by about $\simeq 50\%$ between $l=500$ and
$l=2000$, but they yield compatible results otherwise . The IB estimator does
well for $l>500$ but can be a up to factor of 4 too large for $l<500$.

\subsection{Power spectrum: mass resolution and shot-noise}

\begin{figure}
	{\centering{\epsfysize=7.5cm \epsfbox{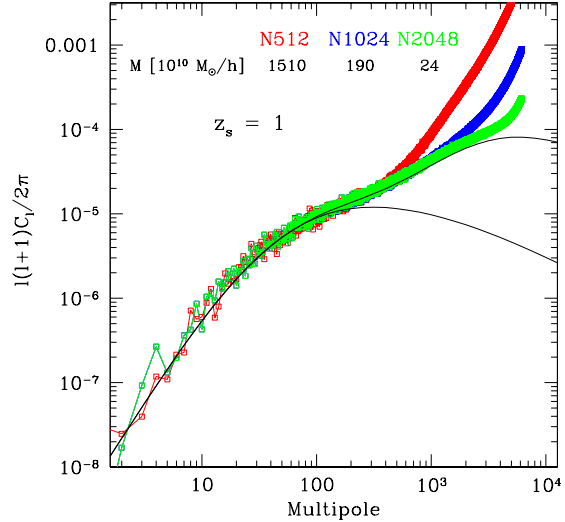}}
	}
\caption{Raw power spectrum estimated from convergence maps in simulations (symbols)
with $z_s=1$ and for different resolutions (shown in Table \ref{table1}). 
Continuous lines show the linear and non-linear halo-fit predictions
which are obtained by replacing $P(k,z)$ in Eq.\ref{clkappa} by the corresponding
3D power spectrum.}
\label{fig:clk}
\end{figure}

Fig.\ref{fig:clk} shows the power spectrum in the convergence maps with
different mass resolution (but same $L_{box}$, see Table
\ref{table1}). There are two different contributions to the effects
shown here. One is the possible difference due to mass resolution and
the other is due to the finite particle density, which results in a different
shot-noise contribution.

\begin{figure}
	{\centering{\epsfysize=7.5cm 
\epsfbox{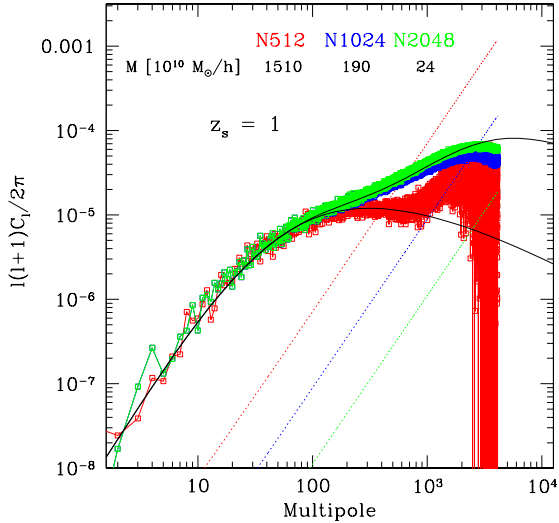}}
	}
\caption{Same as Fig.\ref{fig:clk}, but corrected for shot-noise. The correction
in each case is shown by the dotted lines.}
\label{fig:clks}
\end{figure}

To correct for shot noise we subtract from the measured $C_l$ a term
given by Eq.\ref{clkappasum} with $P(k,z)=1/\bar{N}$, where $\bar{N}$
is the mean galaxy density at each redshift. The corrected spectrum is
shown in Fig.\ref{fig:clks}. The results for the two higher
resolutions agree well up to $l \simeq 1000$, and roughly follow the
halo-fit prescription. The power spectrum estimated from the low resolution map, 
containing $512^3$ particles (N512) and mass resolution 
$\M\simeq 1.5 \times 10^{13} \Msun$), 
closely follows the linear (rather than the non-linear) prediction. 
This could be due to the lack of
small halos and it is also apparent when we compare measurements of the 3D power
spectrum in outputs from the low resolution measured in the low resolution $512^3$ MICE comoving
simulations outputs to the corresponding linear $P(k,z)$ predictions.
The inability of the low resolution simulation to
reproduce non-linear effects reflect the well known fact
(\eg see halofit in \pcite{Smith03}) 
that the power on non-linear scales is
dominated by the internal structure of  halos with mass smaller than
a few times $10^{13} \Msun$, \ie comparable to the particle mass 
of the low resolution simulation.

\subsection{Power spectrum: halo-fit}

\begin{figure}
	{\centering{\epsfysize=8cm \epsfbox{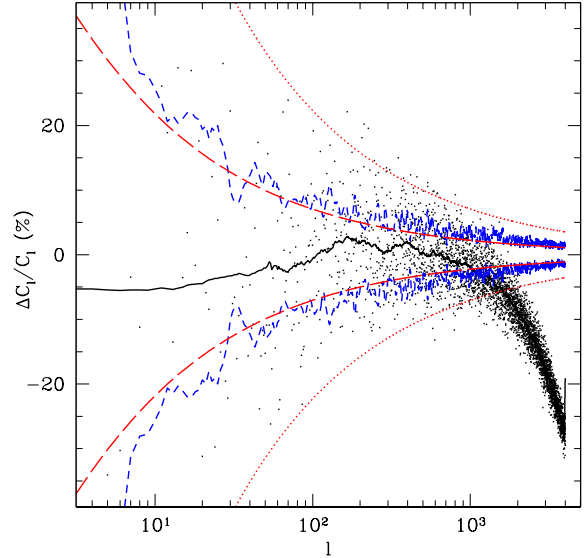}}
	}
\caption{Points show the angular power spectrum of the all-sky convergence maps 
relative to the non-linear halo fit model. The continuous line is a smoothed version
of the dots. The short-dashed line indicates the errors in 10 subsamples with
$10\%$ of the sky. The dotted (long-dashed) lines show the corresponding 
Gaussian error for $10\%$ ($100\%$) of the sky.}
\label{rclk}
\end{figure}

Fig.\ref{rclk} shows the relative difference between the measurements
in the (all-sky) simulations and the halo-fit prediction.  This
prediction is obtained by using the halo-fit model $P(k,z)$ in
Eq.\ref{clkappa}. The dashed lines show the dispersion in 10
subsamples (each $10\%$ of the sky). The halo-fit model only seems to
work within $5\%$ accuracy up to multipoles $l<1000$. Deviations at
smaller scales are significant (up to $30\%$). This is the case even
for a survey which is only $10\%$ of the sky (\ie $f_{sky}=0.1$), shown 
as dashed lines in Fig.\ref{rclk}.  
Note that contrary to Fig.\ref{fig:cle} we have not
binned the data here. This explains the large discrepancy between the
Gaussian errors (dotted lines) and the subsample errors
(short-dashed lines). 
Correlation between bins is strong resulting in smaller diagonal
errorbars, but larger covariance for the subsample errors.
Surprisingly, the Gaussian error prediction for the all-sky
simulations (long dashed lines) is similar to the measured diagonal error in sub-samples
of $f_{sky}=0.1$ (see also \pcite{Cabre07} for a related discussion
in cross-correlation analyses).

\section{Non-Gaussianity and Projection effects}

\subsection{Moments}

It has now been well established that the $p$-order cumulants
of the projected local density field $\mg\kappa^p\md_c$ are expected
to behave as
\beq{eq:moments}
\mg\kappa^p\md_c=S_p\,\mg\kappa^2\md^{p-2}
\eeq
with $p=3,4...$, on large scales (see \pcite{Bernardeau02} and references therein). 
The $S_p$ parameters, quantify the departure from Gaussian behaviour,
and the variance $\mg\kappa^2\md_c$ can be obtained from the power
spectrum above. The measurement of the gravitational weak shear induced by the 
large scale structures in deep
galaxy catalogs will reveal this correlation properties
of the projected mass, at the level of the two-point function
(Blandford et al. 1991, Miralda-Escud\'e, 1991, Villumsen 1996,
Jain \& Seljak 1996, Kaiser 1995) or for higher
orders (Bernardeau, van Wearbeke \& Mellier 1996, Gazta\~naga \& Bernardeau 1998).

\begin{figure}
	{\centering{\epsfysize=8cm \epsfbox{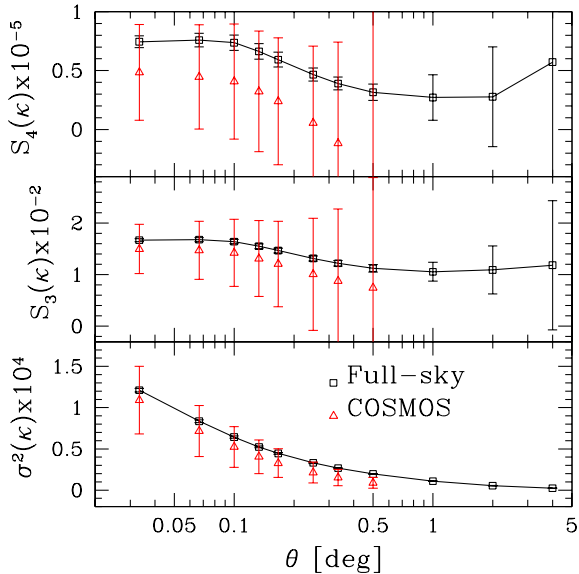}}
	}
\caption{Squares show the variance (bottom), skewness (middle) 
and kurtosis (top) measured in the all-sky convergence map with $z_s=1$.
Triangles show that the mean in $10^4$ small subsamples of 1.6 sqr.degrees
that match the size of the COSMOS HST field is biased with respect the result using all the sky.
The larger errorbars represent the 1-sigma variance in the COSMOS HST subsamples.
The smaller errorbars correspond to $f_{sky}=0.1$.}
\label{fig:moments}
\end{figure}

Figure \ref{fig:moments} shows the above moments for the convergence maps,
smoothed with a Gaussian window of size $\theta$.  Results are in good
agreement with the analytic predictions mentioned above. More details of this
comparison will be given elsewhere.  Here we just want to stress that
the maps are strongly non-Gaussian and we want to focus on the error
estimation.

The smaller errorbars in Figure \ref{fig:moments} correspond to
$f_{sky}=0.1$, while the larger errorbars correspond to a field size
of 1.6 square degrees, comparable to the HST COSMOS field
\cite{COSMOS}. Squares correspond to the mean in the all-sky map,
which agrees very well with the mean of the 10 $f_{sky}=0.1$
subsamples.  This is not the case for the mean in the ($10^4$)
COSMOS-sized subsamples, which is severely biased. 
We find a $10\%$ bias for the variance and skewness at $2^{\prime}$,
and this relative bias increases by a factor of ~2 
for $10^{\prime}$ and a factor of ~4 for $30^{\prime}$ scales.
This sampling bias is common when one has large fluctuations at the scale of the survey
(eg \pcite{HG98}), as is the case here (\eg see also Fig.\ref{kmaps} 
for a visual impression of this effect).

\subsection{Mass reconstruction}

We could in principle use the weak lensing signal to map the 3D mass
distribution in the universe (\eg \pcite{COSMOS}) or we can use it for the more
modest task of calibrating the mass of known clusters with the idea of
estimating the cluster mass function (see \eg \pcite{WWM02}). Here we
want to use the large statistics in our huge volume simulation to 
study how important projection effects could be for this mass
calibration.

At each sky pixel $i$ we find the
redshift onion shell $j$ where the contribution to $\kappa(i)$ in
Eq.\ref{kappasum} is maximum. This produces an array of 3D pixels
$(i,j)$ which are potential sites for large overdensities 
(\eg those one can typically associate to clusters in galaxy surveys)
which we want to use for mass reconstruction. We will assume that both
$j$ and $z_s$ are known and we will use the total convergence
$\kappa(i)$ to estimate the overdensity at $(i,j)$ as: \beq{eq:deltak}
\delta(\kappa) \simeq ~\kappa(i)~{{r_s a_j}\over{(r_s-r_j)r_j dr_j}}
\eeq This procedure mimics a simple linear mass reconstruction method
which assumes that the measured convergence $\kappa(i)$ is dominated
by the overdensity of pixel $(i,j)$ (the maximum along the
line-of-sight).

\begin{figure}
	{\centering{\epsfysize=8cm \epsfbox{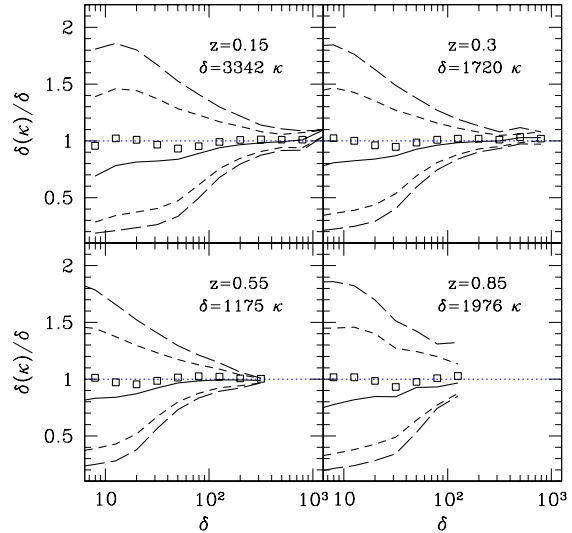}}
	}
\caption{Squares show the mean ratio of reconstructed versus true density fluctuations
in the convergence maps. Continuous lines represents the median of
the distribution ($50\%$ percentile).
The short-dashed (long-dashed) line show $25\%$ ($16\%$) and 
$75\%$ ($84\%$) percentiles. Each panel corresponds to a different
redshift, as labeled in the plots.}
\label{fig:pro}
\end{figure}

Square symbols in Fig.\ref{fig:pro} show the mean ratio between the
reconstructed $\delta(\kappa)$, given by Eq.\ref{eq:deltak} and the
true mean fluctuation $\delta(i,j)$ over all pixel positions $i$ in the
simulation, as a function of the true fluctuation.  Each panel
corresponds to a different redshift (\ie a different value of $j$ in
Eq.\ref{eq:deltak}). The long-dashed lines correspond to the 1-$\sigma$
scatter in the reconstruction. As can be seen in the figure, the mean
reconstruction is basically unbiased but there is quite a large scatter which
increases as we decrease the size of the true fluctuation.

The top panel in Fig.\ref{fig:massf} shows cumulative histograms of
the above reconstruction, where we have converted the fluctuations
$\delta$ into mass using $M=(1+\delta)\bar{\rho} dV$, where
$\bar{\rho}$ is the mean pixel density and $dV$ is the pixel
volume. Note that the distribution is not Gaussian.  These results
agree, at least in a qualitative way, with more detailed studies over
smaller simulations (\eg \pcite{WWM02}).

\subsection{Implications for the mass function}

\begin{figure}
	{\centering{\epsfysize=8cm \epsfbox{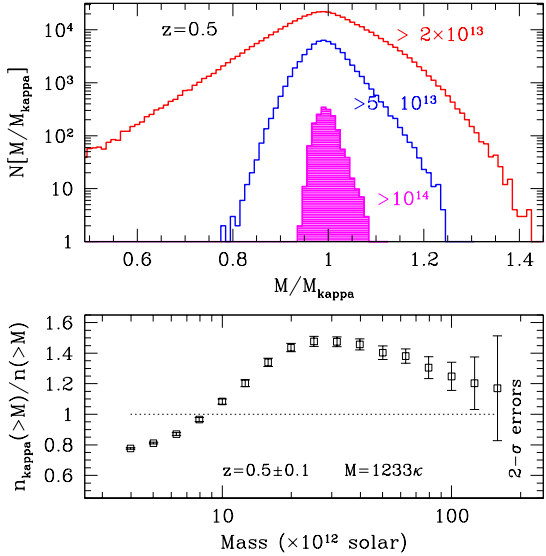}}
	}
\caption{Top: Histogram showing the ratio $M/M_{kappa}$, of true versus
recovered convergence mass, for pixels
in the simulation with true mass $\M$ above $\times 10^{13} \Msun$ (outer histogram),
$5 \times 10^{13}\Msun$ (middle histogram) and $10^{14}\Msun$ (inner histogram). 
The mean recovered mass is not biased, but the distribution is quite broad and
non-Gaussian. Bottom: implication of the above histograms for the recovered
mass function. The ratio of the recovered over the true cumulative number
of pixels above a given mass is shown as a function of the true mass.
Errorbars represent 2-$\sigma$ rms dispersion when we split the simulation in 
10 pieces of $f_{sky}=0.1$ each.
}
\label{fig:massf}
\end{figure}

The large scatter in the above mass reconstruction has important
implications for estimating the (cluster) mass function.  This is
illustrated by the bottom panel of Fig.\ref{fig:massf}, which shows
the ratio of the recovered versus the true cumulative mass function at
$z=0.5 \pm 0.1$. We chose this redshift because it is effectively where the convergence 
window function is maximum for sources at  $z_s=1$. We also show 2-$\sigma$
errorbars from the variance in 10 subsamples with $f_{sky}=0.1$. As
can be seen in the figure, there is a significant bias, given the
errorbars, in the mass function.  Deviations can be as large as $\simeq
50\%$, with an excess density at the high mass end
($\M>10^{13}~\Msun$) and a smaller deficit at the lower mass end.  
This is expected because there is a larger
number density of smaller mass overdensities, which results in a
greater excess of smaller mass objects that scatter into the large
mass bins.

\section{Summary \& Conclusions}   
\label{sec:discuss}   

Radial shells are a natural and convenient decomposition of the data volume 
to exploit large astronomical surveys. Both
because of the limitations in our ability to measure precise redshifts
for all objects in galaxy surveys (especially at $z>1$), and also because it is a natural
way to split the survey data to study evolution or avoid redshift space distortions. 
Here we have presented a new generation of
very large scale N-body simulations developed at the Marenostrum supercomputer using the GADGET-2 code.
We have named them the {\it Marenostrum Institut de Ciencies de l'Espai} or MICE simulations.
In this first paper we have focused on a 
simulation that contains almost $10^{10}$ particles in a
cubical box of $3 \Gpc$ on a side, what delivers good enough mass resolution 
($2.34 \times 10^{11} \Msun$) for studying 
clustering statistics of the large scale structure and
cover a dynamical range of five orders of magnitude: from Gpcs to tens of Kpcs.
This allows us to
sample from the largest (linear) scales to very small (non-linear) structures.

Using this MICE simulation we have built an all-sky
lightcone output that extends to 
high redshift by replicating the parent simulation box $L_{box}$.
We have shown that, thanks to the way we build the lightcone, the effect of repeated 
structures at distances $r > L_{box}$ from the observer is 
effectively negligible in clustering measurements.
In order to mimic the onion-like structure of real data from galaxy surveys
we have compressed the lightcone data into a set of radial shell maps of given redshift resolution. 
These all-sky angular maps have then been
pixelized using the convenient Healpix tesselation with high spatial resolution.
Our approach provides an effectively
lossless method to compress simulated data by a factor $\sim 1000$ 
for arcminute resolution angular maps. 
This allows Terabyte-sized simulations containing tens of
billions of particles to be analyzed in a regular laptop.  

We have presented two main applications of the onion maps
for the study of large-scale sctructure in the universe:

\begin{itemize}
\item  the study of BAOs in angular maps of the dark-matter distribution in the lightcone
\item the construction, for the first time, of an adequate 
all-sky simulated weak lensing map with fine angular resolution.
\end{itemize}

The onion maps we have generated
are large enough to detect the BAO scale with a precision better than
$1\%$. We have presented the angular
power spectrum of the maps (see Figs.\ref{clgg}-\ref{nclggz}) and discussed 
how accurate is the Gaussian
estimation of sampling errors in the presence of non-linear effects.

In Section 3, we have used the onion shells to build a new set of all-sky
lensing maps.  These maps are validated by comparing its power
spectrum and higher orders in the map to theoretical expectations. In
Fig.\ref{fig:clcheck2} we show the relative contribution of each
redshift shell to the total convergence power.  Because we simulate 
the entire celestial sphere we can compare theoretical prescriptions for error
estimation with an error based on the rms dispersion over 
sky patches (subsamples). As summarized in Fig.\ref{fig:cle}, we find 
that Gaussian errors tend to underestimate the true errors in the power spectrum by about
$\simeq 50\%$ at non-linear scales in multipole bins between $l=500$ and $l=2000$, even if we 
use broad bins to minimize covariance between adjacent multipoles.
A comparison of the convergence power spectrum to the analytic halofit predictions 
yields discrepancies of order $30\%$ on
highly non-linear scales, $l>1000$ (see Fig.\ref{fig:clk}-\ref{rclk})  .

We find that current lensing
surveys (such as \pcite{COSMOS}) might yield biased estimates of the
clustering statistics since measurements are subject to large sampling variance.
In other words, these small surveys do not represent a fair sample of
the universe. This has been quantified in the variance and higher-orders 
of the convergence field (see Fig.\ref{fig:moments})
by randomly sampling the all-sky lensing map with $\sim 10^4$ COSMOS-sized surveys.  
For the variance, we estimate a $\simeq 10\%$ bias and $\simeq 40\%$ errors at $2^{\prime}$,
and the bias increases to $25\%$ and $50\%$ at $10^{\prime}$ and $30^{\prime}$ scales.
We also find comparable relative biases for the skewness as a function of scale.

Our estimate of the error is significantly larger than the value reported in \scite{Massey07}
for the COSMOS survey.
The origin of this discrepancy might be the fact that \scite{Massey07} compute 
the variance using {\it subsamples} of the COSMOS data, and thus they do not include
sample variance at the scale of the survey.
Consequently, cosmological constraints on $\sigma_8 \Omega^{0.44}$ based on this estimate of the
variance are expected to be also biased low significantly and the error could 
be underestimated by as much as as a factor of 2.

We have also measured the degree of non-Gaussianity in lensing maps 
induced by non-linear growth.
Higher order moments in the convergence field, shown in
Fig.\ref{fig:moments}, are compatible with theoretical expectations.  In
particular, they match well the amplitudes of the hierarchical scaling
expected from non-linear gravitational clustering \cite{Bernardeau02}.
In the case of weak lensing, these amplitudes are also strongly
dependent on cosmological parameters \pcite{BWM97,GB98}.

Finally, we have presented a mass calibration procedure using lensing maps.
In Fig.\ref{fig:massf} we illustrate how well we expect to recover
mass estimates based on all-sky convergence maps. Upcoming wide surveys, such
as DES \cite{DESWP}, plan to calibrate the cluster mass function using
the weak lensing information. We have shown here that this approach is
a promising tool for calibrating masses, but it needs
to be corrected from systematic biases that arise because of the large scatter
induced by projection effects. Further work needs to be done to check the 
impact of these considerations in specific mass callibration methods, 
such as the one recently presented in \pcite{Johnston07}.

Note added in proof: after our paper was submitted, other papers appeared
presenting applications of N-body simulations for CMB lensing analyses \cite{DB08,Carbone08}.
We note that the method introduced in our paper can be straightforwardly applied to CMB lensing.
We will present this application elsehwere.

\section*{Acknowledgments} 

We would like to thank Anna Cabr\'e, Carlton Baugh, Francis Bernardeau,
Manuel Delfino, Gus Evrard, and Volker Springel, for useful discussions 
at different stages of this work. 
Special thanks for Santi Serrano for his help with visualization tools for the
MICE simulations and Fig.1 of this paper.
We acknowledge support from the MareNostrum supercomputer
(BSC-CNS, www.bsc.es) and Port d'Informaci\'{o} Cient\'{i}fica (PIC, www.pic.es)
where the original simulations where run and stored, the
Spanish Ministerio de Ciencia y   
Tecnologia (MEC), project AYA2006-06341 with EC-FEDER funding
and research project 2005SGR00728 from  Generalitat de Catalunya.
Some of the plots in this paper where produced using CMBview,  
developed by Jamie Portsmouth (www.jportsmouth.com/code/CMBview/cmbview.html).
PF acknowledges support
from the Spanish MEC through a Ramon y Cajal fellowship.
EG would also like to thank the
hospitality of Instituto Nacional de Astrofisica, Optica 
y Electronica (INAOE, Mexico), Galileo Galilei Institute 
for Theoretical Physics (Florence, Italy) and
the Center for Cosmology and Particle Physics (NYU, USA).
MM acknowledges support from the DURSI department of the Generalitat
de Catalunya and the European Social Fund. 
This work was supported by the European Commission's ALFA-II programme
through its funding of the Latin-American European Network for
Astrophysics and Cosmology (LENAC).



\begin{thebibliography}{} 

\bibitem[{Adelman-McCarthy} et al <2006>]{sdssDR4}
Adelman-McCarthy et. al., (The SDSS Collaboration), 2006 Astrophys J. Supp 162, 38-48 

\bibitem[{Annis} et al <2005>]{DESWP}
Annis, J., et al. (Dark Energy Survey white paper) 2005, astro-ph/0510195

\bibitem[{Angulo} et al <2008>]{ABFL08}
Angulo, R.E., Baugh, C.M., Frenk, C.S., Lacey, C.G., 2008, MNRAS, 383, 755

\bibitem[{Bacon} et al. <2001>]{Bacon01}
Bacon D.~J., Refregier A., Clowe D., Ellis R.~S., 2001, MNRAS, 325, 1065 

\bibitem[{Bartelmann \& Schneider}  <2001>]{BS01}
Bartelmann, M \& Schneider, P., 2001, Phys.Rept. 340, 291

\bibitem[{Bernardeau} et al  <1997>]{BWM97}
Bernardeau, F., Waerbeke, L. van  \& Mellier, Y. 1997, A\&A, 322, 1

\bibitem[{Bernardeau, van Waerbeke \& Mellier} <1997>]{BWM97}
Bernardeau, F., van Waerbeke, L., Mellier, Y., 1997 A\&A 322, p1-18

\bibitem[{Beranardeau} et al <2002>]{Bernardeau02}
Bernardeau, F., Colombi, S., Gazta\~naga, E., Scoccimarro, R., 
 2002, Phys. Rept. 367, 1

\bibitem[{Blandford} et al <1991>]{Blandford91} 
Blandford R. D., Saust A. B.,  Brainerd T. G., Villumsen J. V., 1991, MNRAS 251, 600

\bibitem[{Blaizot} et al <2005>]{Blaizot05} 
Blaizot J., Wadadekar Y., Guiderdoni B., Colombi S.~T., Bertin E., Bouchet F.~R., Devriendt J.~E.~G., Hatton S., 2005, MNRAS, 360, 159 

\bibitem[{Bond} et al  <1996>]{BKP96}
Bond, J.R., Kofman, L., Pogosyan, D., 1996, Nature, 380, 603

\bibitem[{Cabre} et al  <2007>]{Cabre07}
Cabre, A., Fosalba, P., Gazta\~naga. E., Manera, M., 2007,
MNRAS, 381, 1347

\bibitem[{Carbone et al.} <2008>]{Carbone08} 
Carbone C., Springel V., Baccigalupi C., Bartelmann M., Matarrese S., 2008, MNRAS, 814 

\bibitem[{Cooray \& Hu}  <2000>]{CH02}
Cooray, A.  \& Hu ,W.,  2002, ApJ, 574, 19

\bibitem[{Das \& Bode} <2008>]{DB08} 
Das S., Bode P., 2008, ApJ, 682, 1 

\bibitem[{Evrard et al} <2002>]{Evrard02}
Evrard, A.E., et al, 2002, ApJ, 573, 7

\bibitem[{Forero-Romero} et al <2007>]{FR07} 
Forero-Romero J.~E., Blaizot J., Devriendt J., van Waerbeke L., Guiderdoni B., 2007, MNRAS, 379, 1507 

\bibitem[{Fosalba} \& {Szapudi}  <2004>]{Fosalba04}
Fosalba, P, \& Szapudi,I., 2004, ApJ 617, L95

\bibitem[{Gazta\~naga \& Bernardeau}  <1998>]{GB98}
Gazta\~naga, E, \& Bernardeau, F., 1998, A\&A 331, 829

\bibitem[{G{\'o}rski} et al <1998>]{GHW99} 
G{\'o}rski, K.~M., Hivon, E., \& Wandelt, B.~D. 1999, in Proc. MPA-ESO Conf.,
Evolution of Large-Scale Structure: From Recombination to Garching, p.37, Ed. A.J.Banday, R.K.Seth, \& L.A.N. da Costa (Enschede: PrintPartners Ipskamp) 

\bibitem[{Hui \& Gazta{\~n}aga} <1999>]{HG98}  
Hui L., Gazta{\~n}aga E., 1999,  ApJ, 519, 622 


\bibitem[{Jain \& Seljak} <1997>]{JS97}
Jain, B. \& Seljak,U.  1997, ApJ, 484, 560

\bibitem[{Jain} et al <2000>]{JSW00}
Jain,B., Slejak, U. \& S.White 2000, ApJ, 530, 547  

\bibitem[{Jarvis} et al.<2006>]{JJBD06} 
Jarvis M., Jain B., Bernstein G., Dolney D., 2006, ApJ, 644, 71 

\bibitem[{Johnston} et al. <2007>]{Johnston07} 
Johnston, D.E., et al., 2007, arXiv:0709.1159

\bibitem[{Kaiser} <1996>]{Kaiser}
Kaiser N., 1995, ApJ, 439, 1

\bibitem[{Kitzbichler \& White} <2007>]{KW07} 
Kitzbichler M.~G., White S.~D.~M., 2007, MNRAS, 376, 2 

\bibitem[{Miralda-Escud\'e} <1991>]{Miralda91}
Miralda-Escude J., 1991, ApJ 380, 1

\bibitem[{Massey} et al  <2007>]{COSMOS}
Massey, R., et al 2007, Nature, 4445, 286

\bibitem[{Massey} et al  <2007>]{Massey07}
Massey, R., et al 2007, ApJS in press, astro-ph/0701480

\bibitem[{Refregier} et al  <2002>]{RRG02}  
{Refregier}, A., {Rhodes}, J., \& Groth, E., 2002, ApJ 572, L131

\bibitem[{Refregier} <2003>]{Refregier03} 
Refregier A., 2003, ARA\&A, 41, 645 

\bibitem[{Scoccimarro} et al <1999>]{Scocci99}
Scoccimarro R., Zaldarriaga, M., \& Hui, L., ApJ, 527, 1 

\bibitem[{Semboloni} et al <2007>]{Semboloni07}
Semboloni, E. et al, 2007, MNRAS, 375, L6

\bibitem[{Smith} et al  <2003>]{Smith03}
Smith, R.E. et al 2003, MNRAS, 341, 1311

\bibitem[{Springel}  <2005>]{Springel05}
Springel, V., 2005, MNRAS, 364, 1105

\bibitem[{Springel} et al  <2005>]{millenium}
Springel, V. et al 2005, Nature, 435, 639

\bibitem[{Szapudi} et al <2001a>]{spice}
Szapudi,I., 2001, ApJ 548, L115

\bibitem[{Szapudi} et al <2001b>]{spice2}
Szapudi,I., Prunet, S., \& Colombi, S., 2001, ApJ, 561, L11

\bibitem[{Vale \& White}  <2003>]{VW03}
Vale, C. \& White, M., 2003, ApJ, 592, 699

\bibitem[{Villumsen} <1996>]{Vill96} 
Villumsen, J V 1996, MNRAS, 281, 369

\bibitem[{Waerbeke} et al <2000>]{W01}
Waerbeke L.van  et al. 2001, MNRAS, 322, 918

\bibitem[{Wambsganss} et al <2000>]{WCO98}
Wambsganss,J.  Cen, R. \& Ostriker, J.,P., 1998, ApJ, 494, 29 

\bibitem[{White \& Hu}  <2000>]{WH00}
White, M.  \& Hu,W.,  2000, ApJ, 537, 1 

\bibitem[{White, Waerbeke \& Mackey}  <2002>]{WWM02}
White, M., Waerbeke, L. van  \& Mackey, J., 2002, ApJ, 575, 640

\bibitem[{White \& Vale}  <2004>]{WV04}
White, M. \& Vale, C.,  2004, Astrop.Phys, 22, 19





\end{thebibliography}
\end{document}